\documentclass[sigconf]{acmart}
\usepackage{nccmath}
\usepackage{multirow}
\usepackage{caption}
\usepackage{subcaption}
\AtBeginDocument{%
  \providecommand\BibTeX{{%
    \normalfont B\kern-0.5em{\scshape i\kern-0.25em b}\kern-0.8em\TeX}}}

\copyrightyear{2021} 
\acmYear{2021} 
\setcopyright{acmcopyright}\acmConference[WI-IAT '21]{IEEE/WIC/ACM International Conference on Web Intelligence}{December 14--17, 2021}{ESSENDON, VIC, Australia}
\acmBooktitle{IEEE/WIC/ACM International Conference on Web Intelligence (WI-IAT '21), December 14--17, 2021, ESSENDON, VIC, Australia}
\acmPrice{15.00}
\acmDOI{10.1145/3486622.3494002}
\acmISBN{978-1-4503-9115-3/21/12}

\begin{document}

\title[Recommending Multiple Positive Citations via MP-BERT4REC]{Recommending Multiple Positive Citations for Manuscript via Content-Dependent Modeling and Multi-Positive Triplet}

% \titlerunning{Recommending Multiple Positive Citations via MP-BERT4REC}
% \author{Yang Zhang \and Qiang Ma}
% %
% % \authorrunning{Y. Zhang and Q. Ma}
% \aff{Kyoto University, Japan 606-8371, Japan \\
% email{zhang.yang.33z@st.kyoto-u.ac.jp} \\
% email{qiang@i.kyoto-u.ac.jp}}
\author{Yang Zhang}
\email{zhang.yang.33z@st.kyoto-u.ac.jp}
\affiliation{%
  \institution{Graduate School of Informatics, Kyoto University}
  \country{Kyoto, Japan}
  }

\author{Qiang Ma}
\email{qiang@i.kyoto-u.ac.jp}
\affiliation{%
  \institution{Graduate School of Informatics, Kyoto University}
  \country{Kyoto, Japan}
  }


\begin{abstract}
Considering the rapidly increasing number of academic papers,  searching for and citing appropriate references have become a non-trial task during the wiring of papers. Recommending a handful of candidate papers to a manuscript before publication could ease the burden of the authors, and help the reviewers to check the completeness of the cited resources. Conventional approaches on citation recommendation generally consider recommending one ground-truth citation for a query context from an input manuscript, but lack of consideration on co-citation recommendations. However, a piece of context often needs to be supported by two or more co-citation pairs. Here, we propose a novel scientific paper modelling for citation recommendations, namely Multi-Positive BERT Model for Citation Recommendation (MP-BERT4CR), complied with a series of Multi-Positive Triplet objectives to recommend multiple positive citations for a query context. The proposed approach has the following advantages: First, the proposed multi-positive objectives are effective to recommend multiple positive candidates. Second, we adopt noise distributions which are built based on the historical co-citation frequencies, so that MP-BERT4CR is not only effective on recommending high-frequent co-citation pairs; but also the performances on retrieving the low-frequent ones are significantly improved. Third, we propose a dynamic context sampling strategy which captures the ``macro-scoped'' citing intents from a manuscript and empowers the citation embeddings to be content-dependent, which allow the algorithm to further improve the performances. Single and multiple positive recommendation experiments testified that MP-BERT4CR delivered significant improvements. In addition, MP-BERT4CR are also effective in retrieving the full list of co-citations, and historically low-frequent co-citation pairs compared with the prior works.
\end{abstract}

%   <concept>
%       <concept_id>10002951.10003317.10003347.10003350</concept_id>
%       <concept_desc>Information systems~Recommender systems</concept_desc>
%       <concept_significance>500</concept_significance>
%       </concept>
%   <concept>
%       <concept_id>10002951.10003317.10003318.10003323</concept_id>
%       <concept_desc>Information systems~Data encoding and canonicalization</concept_desc>
%       <concept_significance>300</concept_significance>
%       </concept>
%  </ccs2012>

% \ccsdesc[500]{Information systems~Recommender systems}
% \ccsdesc[300]{Information systems~Data encoding and canonicalization}




\maketitle

\section{Introduction}
% When writing an academic paper, one of the most common questions to consider would be ``Which paper should I cite at this point?''. Given the massive number of published papers, it is time consuming to search and read all the relevant papers to a topic. Hence, recommending a handful appropriate papers for the context needed support might be helpful to ease the burden of the authors. A hypothesized application scenario is demonstrated in Figure \ref{fig:application}, where the system takes a query context as input to extract the citing intent of the user, based on which, it provides candidate papers for citing.
Considering the massive amount of academic papers, which is accounted for over 300 million since 2018, and growing at 5\% per year according to \citet{johnson2018stm}. It becomes a challenging task for researchers to search and find appreciate papers for citing when they are writing their own papers, and for the reviewers to check the completeness of the cited resources.

% \begin{figure}[t]
%     \centering
%     \includegraphics[width=0.9\columnwidth]{application.png}
%     \caption{Citation recommendation for (a) context with a singe positive reference, and (b) context with multiple positive references}
%     \label{fig:application}
% \end{figure}

Currently, researchers are generally relying on keyword-based search engines (such as Google Scholar) for searching relevant papers via input keywords. However, it is argued that the input keywords might be over-simplified to reflect the users' searching needs \cite{DBLP:conf/asunam/JiaS17,DBLP:conf/ecir/JiaS18}, and hence they often lead to unsatisfactory recommendations. To appropriately detect the citing intent of the users, a line of studies considered to adapt a collection of seed papers (e.g. the papers previously cited or read by the user) to find the topically relevant articles via collaborative filtering \cite{DBLP:conf/cscw/McNeeACGLRKR02}, random-walk methods \cite{DBLP:conf/webi/GoriP06,DBLP:conf/asunam/KucuktuncSKC13,DBLP:conf/ecir/JiaS18}, or matrix decomposition methods \cite{DBLP:conf/jcdl/CarageaSMG13}. Seed papers could potentially reflect richer information than keywords. However, these methods still lack considerations on extracting the citing intent from the user's manuscript. Later researches utilized the ``local context'' (surrounding words around a target citation) to extract the semantics of the citing intent and then rank the most similar papers via citations embeddings \cite{DBLP:conf/acl/ShiSZZH18,DBLP:conf/coling/ZhangM20}.

Context-based approaches \cite{DBLP:conf/acl/ShiSZZH18,DBLP:conf/coling/ZhangM20} detect citing intents from the sentences needed support, and based on which it finds the best-matched candidate papers. These approaches could be potentially applied in the real world to assist researchers in finding citations during the writing of papers. Nevertheless, context-based methods may still be constrained in four aspects. First, they generally consider recommend one ground-truth citation for the query context. However, it is common that a piece of context cites more than one reference to form co-citations. Second, we aim to improve the performances on retrieving both high and low frequent co-citation pairs by adapting noise frequency distributions which are built based on historical co-citation frequencies. Prior works (\cite{DBLP:journals/jasis/Small73,Gipp2009Citat-31366}) generally have been testified effectively \cite{DBLP:conf/jcdl/KobayashiS018,DBLP:conf/acl/ShiSZZH18} on retrieving the most frequently occurred co-citation pairs in history. However, co-citation pairs that were published in later years, or by non-scientific-elites \cite{DBLP:journals/scientometrics/ParkerAL13} may come with lower occurrences in history. We consider low-frequent co-citation pair could also demonstrate high topic relatedness as studied by \citet{DBLP:journals/jasis/Small73}, and therefore can potentially be utilized to improve recommendation performances. Third,  ``local context'' is helpful to determine the citing intent in a micro-scope \cite{DBLP:conf/acl/ShiSZZH18,DBLP:journals/access/ZhangM20b}. However, we argue that a lack of consideration of the rest of the body content may lead to inaccurate judgments on extracting the topic semantics of the manuscript. Forth, the prior works might be limited to representing the content semantics by adapting the label-based embeddings \cite{DBLP:conf/acl/ShiSZZH18,DBLP:journals/access/ZhangM20b} that do not contain content knowledge, or content embeddings generated solely from abstracts \cite{DBLP:conf/acl/CohanFBDW20}.

% \footnote{According to the researches \cite{DBLP:journals/scientometrics/ParkerAL13}, citations tend to approximate the ``20/80'' phenomenon, meaning that 20\% of papers contributed 80\% of citations. The highly cited papers are testified to follow some specific characteristics: (1) highly productive scholars; (2) top venues; (3) collaborated articles; (4) authors from North Americans and Western Europeans. These researchers are defined as the ``scientific elites''.}

\begin{figure}
    \centering
    \includegraphics[width=0.8\columnwidth]{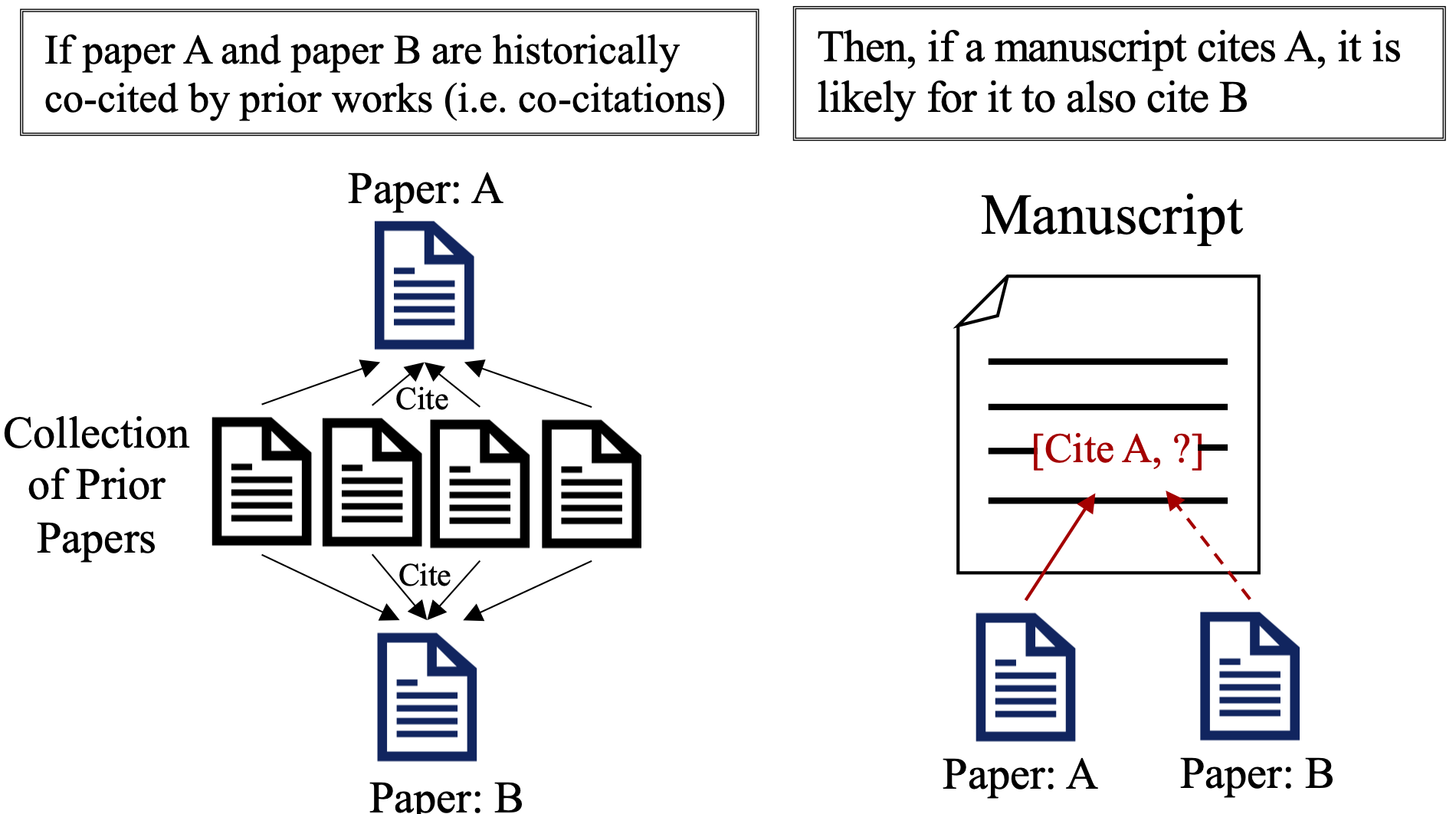}
    \caption{Recommending multiple positive citations by adapting historical co-citation frequencies}
    \label{fig:historical_cocit}
\end{figure}

To this end, we propose a \textbf{M}ulti-\textbf{P}ositive \textbf{BERT} Model for \textbf{C}itation \textbf{R}ecommendation (MP-BERT4CR), which is a content model for citation recommendations coupled with multi-positive triplet objectives. It has the following advantages: First, the proposed multi-positive triplet objective functions allow the algorithm to be optimized for recommending both single and multiple positive candidates, which can effectively find multiple positive candidates than conventional triplet objective. Second, we leverage the historical co-citation frequencies to generate positive samples. The underlying logic is: if two (or more) citations historically appeared as co-citations in the past, and one of them is cited by a manuscript at the present, then it is likely that other co-cited papers should be cited, as illustrated in Figure \ref{fig:historical_cocit}. We build noise distributions based on historical frequencies to draw co-citation samples, which is effective in retrieving both high and low frequent co-citations in history. Third, a dynamic context sampling strategy is proposed to extract the citing intents in a macro-scope by extracting the topic semantics from the remaining part of the body content of a manuscript, and producing citation embeddings carrying content semantics from candidate papers. Recommendations made by matching the content semantics of candidate papers with the ``macro-scoped'' citing intents produced superior performances compared with the baselines.

\raggedbottom

% \begin{itemize}

%     \item 
    
%     \item 
    
%     \item 
% \end{itemize}
 
% \noindent In addition, the proposed framework could also be adapted for recommendations of other types of texts.

We conduct experiments on single and multiple citation recommendations to evaluate MP-BERT4CR with the conventional baseline models in the field on four datasets. The results revealed that the MP-BERT4CR had provided significant improvements for both single and multiple positive recommendations compared to the baselines.

\vspace{-1mm}

\section{Related Work}

\subsection{Citation Recommendation}
Citation recommendation denotes the task of finding relevant papers based on an input query. A line of studies use a collection of seed papers as the input query to find topically relevant papers via collaborative filtering \cite{DBLP:conf/cscw/McNeeACGLRKR02}, random-walk methods \cite{DBLP:conf/webi/GoriP06,DBLP:conf/asunam/KucuktuncSKC13,DBLP:conf/ecir/JiaS18}, or matrix decomposition methods \cite{DBLP:conf/jcdl/CarageaSMG13}. These methods could be helpful during the surveying of a topic at an early stage. However, when applying to assist the writing of a paper, they generally lack considerations on detecting the citing intents of users. Context-based methods \cite{DBLP:conf/acl/ShiSZZH18,DBLP:conf/coling/ZhangM20} aimed to extract citing intent via an input local context (surrounding words around a target citation), and hence find the most relevant papers based on the detected citing intent. Nevertheless, their methods still leave room for further improvements. First, previous methods only consider recommending one positive citation to a context, which may constraint the usability; second, local context may only reveal the ``micro-scoped'' citing intent which could not infer the topic in a macro-scope of the manuscript, and therefore may lead to inaccurate judgments on extracting the true citing intent of users; third, these method adapted label-based embeddings without content semantics, which may further constraint the performances.

% \subsection{Co-citation Analysis}

% Co-citation relationship is initially defined as two (or more) prior works are cited together by the later literature \cite{DBLP:journals/jasis/Small73}. According to the qualitative analyses by \citet{DBLP:journals/jasis/Small73}, the majority of the co-cited papers come with direct citation relations and bibliographic coupling relations. Co-citations also demonstrated strong topic relatedness. \citet{Gipp2009Citat-31366} testified the strong strength between co-citations through human examiners. \citet{DBLP:conf/jcdl/KobayashiS018} conducted recommendation tasks for co-citation recommendations, however the adapted datasets are relatively small (about 20,000 papers) and they did not comprehensively compared with best-performed baselines and tasks for non-co-citations. In this paper, we leverage the information of strong topic relevance carried by co-citation pairs pointed from the prior works, to improve the recommendation performances by building noise distributions and optimizing via multi-positive objective functions.
\vspace{-1mm}

\subsection{Document Embedding}
Document Embedding refers to the studies on representing documents as continuous vectors. The early approaches, Word2Vec \cite{DBLP:journals/corr/abs-1301-3781} and Doc2Vec \cite{DBLP:conf/icml/LeM14} were proposed to learn word embeddings by preserving the contextual information via DNN-like networks. However, Word2Vec and Doc2Vec generally treated the input documents as ``plain texts'' which may lead to information loss issues. Later approaches were considered to fine-tune with specific information from academic papers, such as hyperlinks \cite{DBLP:conf/acl/ShiSZZH18}, and section headers and word-wise relations in the local context \cite{DBLP:conf/coling/ZhangM20}. The recent language modelling models such as BERT \cite{DBLP:conf/naacl/DevlinCLT19}, SciBERT \cite{beltagy-etal-2019-scibert} are effective on multiple NLP tasks. In this study, we further extend the language modeling for citation recommendations, considering multiple positive candidates. 

\vspace{-1mm}

\section{Problem Definition}

We adapt part of terminologies and definitions for academic papers, citation relationships, and citation recommendation following the past studies \cite{DBLP:conf/acl/ShiSZZH18,DBLP:conf/coling/ZhangM20}.

\subsection{Terminologies and definitions}

Let $\mathit{w} \in \mathit{W}$ represent a word from a vocabulary set $\mathit{W}$, and $\mathcal{H}_i$ represent the $\mathit{i}$-th paper from a text collection $\mathbb{H}$.

% The content words appearing in $\mathcal{H}_i$ are denoted as collections of words $\mathit{\hat{W}} \subseteq \mathit{W}$. The document IDs of the cited papers in $\mathcal{H}_i$ are denoted as a collection $\mathit{\hat{D}^c} \subset \mathit{D}$.

\begin{definition}[Academic Paper]\label{def:paper}
The textual information of paper $\mathcal{H}_i$ is represented as its content words, and cited papers, i.e., $\mathcal{H}_i := \mathit{\hat{W}_i} \cup \mathbb{\hat{H}}_i^c$, where $\mathit{\hat{W}} \subseteq \mathit{W}$, and $\mathbb{\hat{H}}_i^c \subset \mathbb{H}$. 
\end{definition}

\begin{definition}[Citation Relationship]\label{def:cit}
Given a source paper, $\mathcal{H}_s:= \mathit{\hat{W}}_s \cup \mathbb{\hat{H}}_s^c$, cites a target paper, $\mathcal{H}_t :=  \mathit{\hat{W}_t} \cup \mathbb{\hat{H}}_t^c$, where $\mathcal{H}_t \in \mathbb{\hat{H}}_s^c$. A citation relationship is constructed and expressed using a tuple, which is composed content words of the source papers, and the target paper, i.e., $\mathcal{R} := \langle  \mathit{\hat{W}_s},  \mathit{\hat{W}_t} \rangle$.
\end{definition}

% \begin{definition}[Embedding Model] \label{def:embed}
% An embedding model is defined as a function $\varmathbb{E}$ which represents a paper $\mathcal{H}$ into a $d$-dimensional vector, i.e. $\mathbf{d} = \varmathbb{E}(\mathcal{H})$.
% \end{definition}

\vspace{-2mm}

\subsection{Task Definition on Citation Recommendation}
Given a query paper $\mathcal{H}_s$, a collection of papers $\mathbb{H}$, and an embedding model $\varmathbb{E}$, the task is defined to find top $k$ ranked papers $\{ \mathcal{H}_1, \mathcal{H}_2, ..., \mathcal{H}_k \}$ based on the geometric distances between embeddings of $\mathbb{H}$, i.e., $\mathbf{D} = \varmathbb{E}(\mathbb{H})$ and the embedding of $\mathcal{H}_s$, i.e., $\mathbf{d}_s = \varmathbb{E}(\mathcal{H}_s)$.

\section{MP-BERT4CR}
The algorithm firstly composes the content words of the input papers into a hierarchical structural for encoding. Given a paper defined in Definition \ref{def:paper}, where $\mathcal{H}_i := \mathit{\hat{W}_i} \cup \mathit{\hat{D}_i^c}$, the content words, $\mathit{\hat{W}_i}$ are categorized into sentences, i,e, $\mathit{\hat{W}^{categorized}_i} := \{\mathcal{S}^i_1, \mathcal{S}^i_2,...,\mathcal{S}^i_{|\mathcal{D}|} \}$, where $\mathcal{S}^i_j := \{ w^j_1, w^j_2,...,w^j_{|\mathcal{S}_j|} \}$. At the end of each sentence, a special \textit{end-of-sentence} (EOS) token is added.

We adapt the hierarchical transformer \cite{DBLP:conf/acl/ZhangWZ19} to pre-train the model. However,faewfwfwe replace the EOS pooling to MEAN pooling at the sentence-level encoder from the original model.
% The algorithm then constructs the paper-level content modelling via pre-training, and then fine-tunes for recommendation modelling.

\begin{figure*}[tb]
    \centering
    \begin{subfigure}[b]{0.26\textwidth}
    \includegraphics[width=\textwidth]{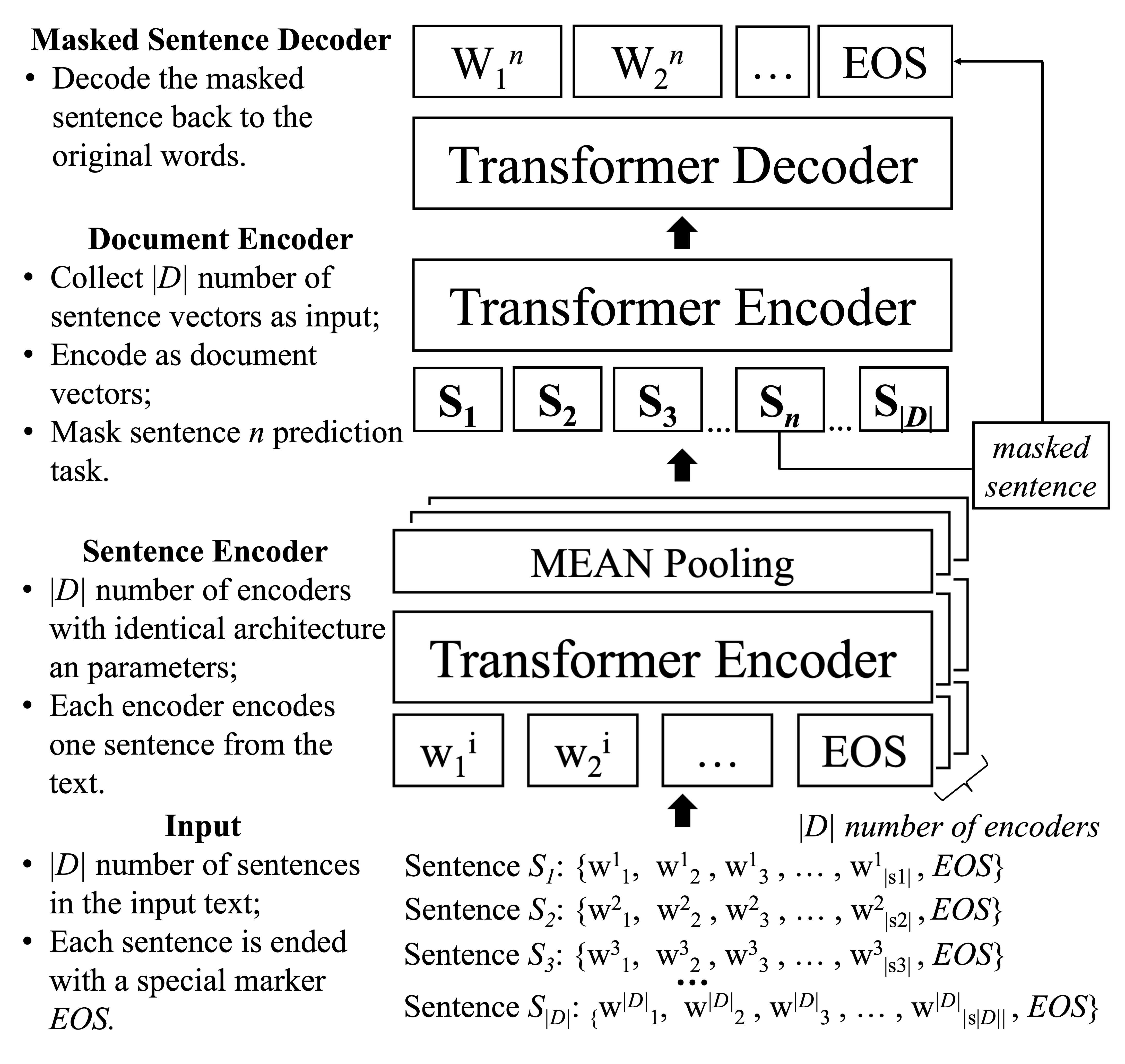}
    \caption{Pre-training model}
    \end{subfigure}
    \begin{subfigure}[b]{0.5\textwidth}
    \includegraphics[width=\textwidth]{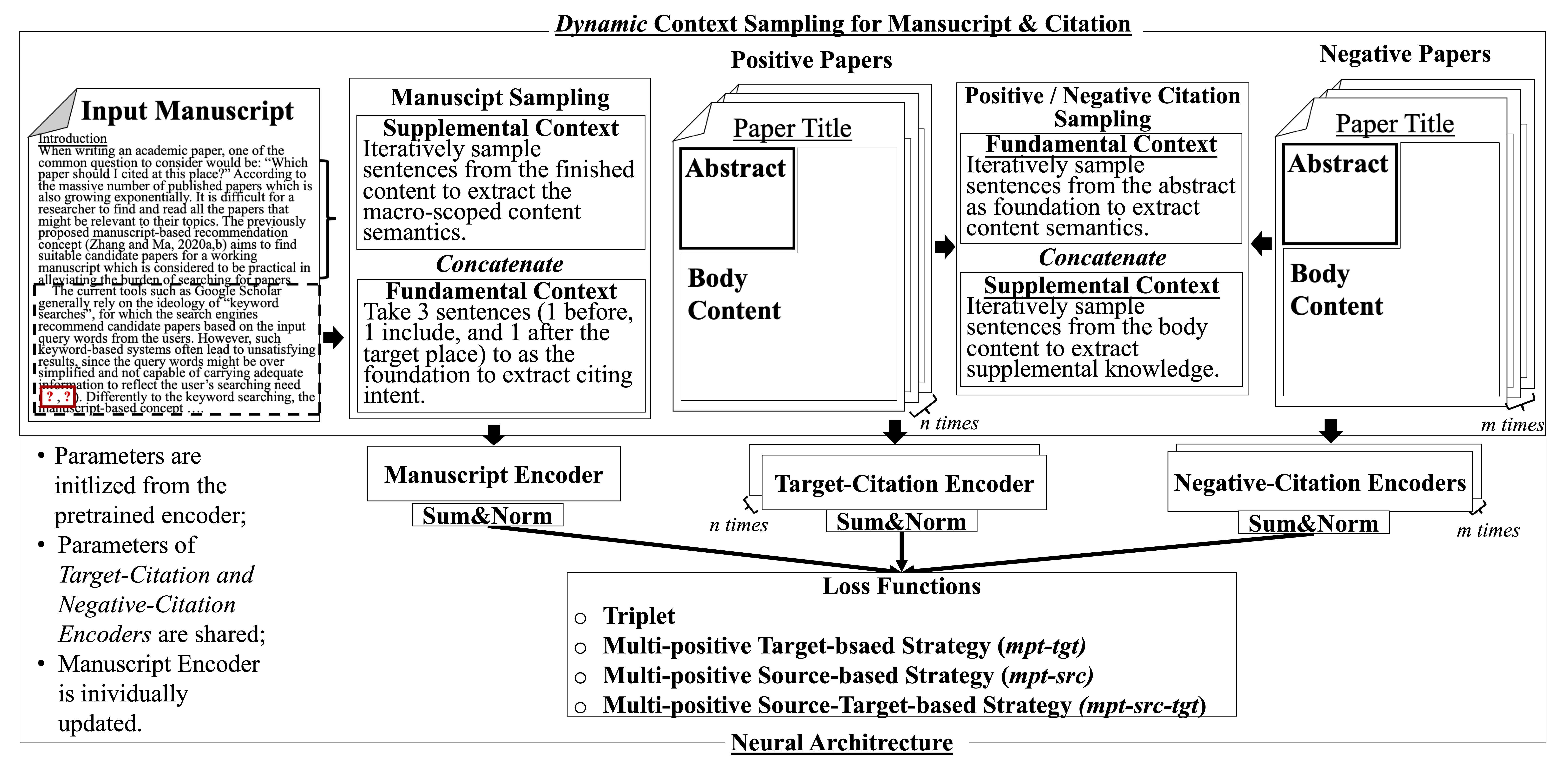}
    \caption{Fine-tuning model}
    \end{subfigure}
    \caption{Overview of MP-BERT4CR}
    \label{fig:model}
\end{figure*}

\subsection{Fine-tuning for Recommendation Modelling}

Fine-tuning is conducted after pre-training to maximize the similarity between the citing intent of an input manuscript and the content semantics of its ground-truth citations. 

We first conduct dynamic context sampling for two purposes: 1) to extract the citing intent from the input manuscript from both of ``micro-scoped'' and ``macro-scoped'' view; 2) to generate content-depend embeddings carrying content semantics for the citations.

% Previous studies \cite{DBLP:conf/acl/ShiSZZH18,DBLP:conf/coling/ZhangM20} generally adopted the stabilized sampling strategy, which uses the local context (the adjacent words around a citation) from a manuscript to predict the label (i.e., the document ID) of the target citation. We consider the citing intent extracted from the local context might merely reflect ``micro-scped'' citing intent, where the topic semantics in a broader context might be lost; in addition, the label embeddings of the citations do not carry content semantics from the contents. Here, we proposed to extract both of the ``micro-scoped'' citing intent from the local context, and ``macro-scoped'' citing intent from the finished content of the manuscript to help determine the citing intent in a comprehensive manner. More over, we also utilize the content from the citations to generate content-dependent embeddings, the recommendations can be made according to the content knowledge. To alleviate the GPU memory bottle-neck, we designed \textbf{dynamic context sampling strategies} to sample essential contexts from a manuscript and its citations, to help capture the macro-scoped citing intent from the manuscript and content semantics from citations. 

And then, we adapt a triplet-encoder neural architecture \cite{DBLP:conf/emnlp/ReimersG19,DBLP:conf/cvpr/SchroffKP15} coupled with multi-positive triplet objectives to encode the sampled contexts, for optimizing the algorithm for single and multiple recommendations.

\subsubsection{Dynamic Context Sampling} \label{sec:sampling}
It aims to extract essential contexts from the manuscript regarding a predicting location for detecting the micro and macro scoped citing intent, as well as capturing content semantics of the cited papers and other positive samples. The sampled contexts involve two components: a \textbf{fundamental context} functions as the backbone for inferring the citing intent or content semantics, and a \textbf{supplemental context} aims to provide additional knowledge on the topic semantics of the paper. The detail of the sampling strategy is illustrated in Figure \ref{fig:model}(b).

As illustrated in Figure \ref{fig:model}(b), for a manuscript, the \textbf{fundamental context} is defined to include three sentences: the sentences include the target citations, the sentence before the target citation, and the sentence after it, which is functionally the same as the local context defined in prior works \cite{DBLP:conf/acl/ShiSZZH18,DBLP:conf/coling/ZhangM20} to infer the citing intent in a micro-scope. The \textbf{supplemental context} is defined as a pre-set number of sentences randomly selected from the finished content (the sentences appearing before the citation excluding the base context) to infer the overall topic of the manuscript, and thus to help determine the citing intent in a macro-scope.

For a cited paper, the \textbf{fundamental context} is defined to be the sentences from the abstract, which works as the backbone for inferring the content semantics. The \textbf{supplemental context} is defined as a pre-set number of randomly selected sentences from the body content to provide additional information on the semantics.

The default settings of the dynamic context sampling strategy were as follows: For the manuscript, we set the total number of sentences to 30, which includes the fundamental context composed of three sentences (the sentence including the target citation, the sentence before it, and the sentence after). Supplemental contexts include 27 sentences, which are randomly selected from the content before the base context. For a citation paper, we set the total number of sampling contexts to 40, which includes 10 sentences for the fundamental context (as it is found that the abstract averagely contain 10 sentences from our datasets) and 30 for the supplemental context (randomly selected from the body content). The number of the selected sentence was chosen according to the maximum memory of our GPUs. In addition, 40 sentences cover 10\% of the contents from our dataset (papers come with 317 sentences overage). We presume that the abstract and 10\% (maximumly 20\% for 2 iterations of fine-tuning) of the body content are sufficient to capture the content semantics.

\vspace{0.5em}

\subsubsection{Neural Architecture} It comprises a \textit{manuscript encoder} for encoding the sampled context from the manuscript regarding a ground-truth citation, a \textit{citation encoder} for encoding the sampled context from the ground-truth citation and other positively sampled citations, and a pre-defined number of \textit{negative-citation encoders} for encoding the negatively sampled citations (Figure \ref{fig:model}(b)). The architecture was inspired by the prior works \cite{DBLP:conf/emnlp/ReimersG19,DBLP:conf/cvpr/SchroffKP15}, where the three encoders are identical to the encoders proposed in the pre-training model of MP-BERT4CR. The parameters of the three encoders are initialized from the pre-trained encoder. During training, the parameters of the \textit{citation encoder} and \textit{negative-citation encoders} are completely shared, whereas the \textit{manuscript encoder} is individually updated. Additionally, a sum pooling layer and a normalization layer \cite{DBLP:journals/corr/BaKH16} are added: the former summarizes the sentence vectors into a document vector, and the latter facilitates the convergence of the objective functions.

\subsubsection{Multi-Positive Sampling}
For the query context with multiple ground-truth citations, in addition to the target citation, we sample a pre-defined number of citations according to their historical co-citation frequencies with the target citation as the additional positive samples. Noise distribution is built based on the historical frequencies of co-citation pairs are built, so that both high frequent and low frequent pairs are assigned a probability to be drawn. High frequent pairs are assigned a relatively higher probability, since it is likely for a pair to be co-cited again, if they are frequently cited before. The probability for low frequent pairs can be adjusted by the power value, which is set to $\frac{3}{4}$ for a default value in Equation \ref{eqn:pos_sample}.

Specially, given $\mathcal{H}_{t*}$ as the target ground-truth citation for prediction, and $\mathbb{H}_{p}$ for the full list of co-citations, the algorithm samples $n$ number (a pre-defined value) of positive citations from the paper collection $\mathbb{H}_d$ via the noise probability distribution \cite{DBLP:journals/corr/abs-1301-3781}: 
% Specially, given $\mathbb{H}_t:=\{\mathcal{H}_{t1}, \mathcal{H}_{t2}, ..., \mathcal{H}_{tk}\}$ as the ground-truth citations for a piece of context, the algorithm samples $n$ number (a pre-defined value) of positive citations $\{\mathcal{H}_{ti} | i \leq n\}$ from the paper collection $\mathbb{H}_t$. Assume the target citation at the current iteration of training is $\mathcal{H}_{t*}$ from $\mathbb{H}_t$, then the noise distribution \cite{DBLP:journals/corr/abs-1301-3781} is constructed:

% based on the co-citation frequencies, $ \mathbb{H}_p := \{ \mathcal{H}_{ti} | 1 \leq ti \leq tk, ti \neq t* \} $ and $\mathcal{H}_{t*}$ with all the papers in the dataset:

\begin{equation}
    \begin{medsize}
    \mathbf{p}(\mathcal{H}_{ti} \in \mathbb{H}_{p}) = \frac{\mathbf{frequency(\mathcal{H}_{t*},\mathcal{H}_{ti})}^{\frac{3}{4}}}{\sum_{\mathcal{H}_{tj} \in \mathbb{H}_d}\mathbf{frequency(\mathcal{H}_{t*},\mathcal{H}_{tj})}^{\frac{3}{4}}},
    \end{medsize}
    \label{eqn:pos_sample}
\end{equation}

\begin{figure}[t]
    \centering
    \includegraphics[width=0.9\columnwidth]{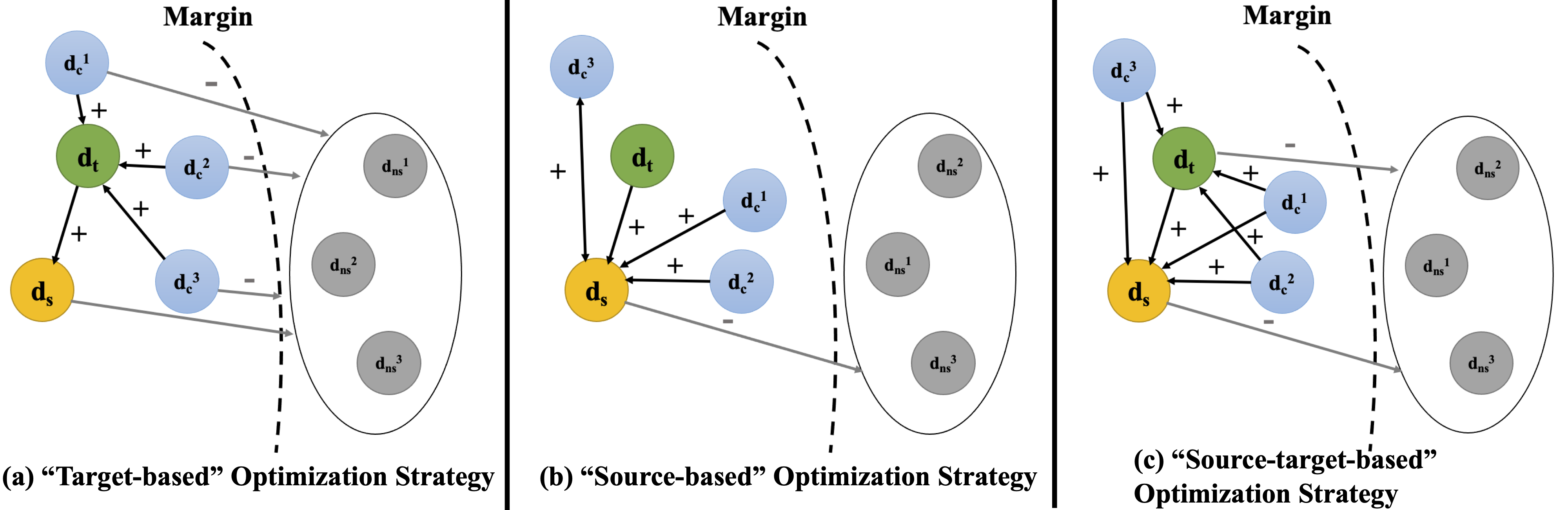}
    \caption{Illustration of Optimization Strategies}
    \label{fig:optim_strategies}
    \vspace{-1em}
\end{figure}
\raggedbottom

\noindent where $frequency$ denotes the count of the two input papers being appeared as co-citations from the dataset. The number of positive samples is set to be 3, as it is found that the number of co-citations with greater than 3 items are neglectable small in our datasets.

%$\mathbb{H}_{pn}$ represents the sampled positive candidates.

% A pre-defined $n$ number of citations are drawn from the probability $\mathbf{p}(\mathcal{H}_{ti} \in \mathbb{H}_{p})$ as the positive samples, noted as $\mathbb{H}_{pn}$.

\subsubsection{Negative Sampling}

A negative sampling strategy is adapted to pick the papers which are not cited by the input context as the targets for similarity minimization. The objective is that the irrelevant papers should not appear in the top recommendation list.

The negative citations are sampled based on their occurrences as citations. The underlying intuition is that if a paper is frequently cited by other papers, however, it is not cited by the input context, then it would be drawn more frequently for similarity minimization between it and the input context. Similar to positive sampling, a noise distribution is constructed based on their occurrences as citations:

\begin{equation}
    \begin{medsize}
    \mathbf{p}(\mathcal{H}_i \in \mathbb{H}) = \frac{\mathbf{count(\mathcal{H}_i)}^{\frac{3}{4}}}{\sum_{\mathcal{H}_j \in \mathbb{H}}\mathbf{count(\mathcal{H}_j)}^{\frac{3}{4}}},
    \end{medsize}
    \label{eqn:neg_sample}
\end{equation}

\noindent where $\mathbb{H}$ denotes all the papers in the dataset except the positive citations; and $count$ denotes the number of occurrences as citations of a paper. A pre-defined $m$ number of papers are picked from the distribution as negative samples, noted as $\mathbb{H}_{ns}$. We set $m$ to be 4 in this study.

\subsubsection{Multi-Positive Triplet Objectives} They are designed to optimize the algorithm for recommending multiple positive citations. 

Suppose a piece of context in the manuscript,$\mathcal{H}_s$ containing the ground-truth citation, $\mathcal{H}_t$, along with $N$ number of co-citations $\mathcal{H}_t$, i.e., $\mathcal{H}_{c}^1,...,\mathcal{H}_c^N$. The algorithm retrieves $n$ positive samples $\{\mathcal{H}_c^n | 1 \leq n \leq N \}$ from Equation \ref{eqn:pos_sample}, and $m$ negative samples $\{\mathcal{H}_{ns}^m | 1 \leq m \}$ from Equation \ref{eqn:neg_sample}. 

% Then the algorithm conducts dynamic manuscript sampling for $\mathcal{H}_s$, and dynamic citation sampling for the target citation, positive and negative sampled citations as explained in section \ref{sec:sampling}, which are then encoded as $\mathbf{d}_s$ via the manuscript encoder; and $\mathbf{d}_t$, $\{\mathbf{d}_c^n | 1 \leq c \leq n \}$, and $\{ \mathbf{d}_{ns}^m | 1 \leq m \}$ via the citation encoder.

We propose multiple positive objectives considering the multiple positive samplings by modifying the original triplet-encoder \cite{DBLP:conf/emnlp/ReimersG19,DBLP:conf/cvpr/SchroffKP15} which originally considers one target and one negative sample:
\begin{equation}
    \begin{medsize}
        \max(||\mathbf{d}_s - \mathbf{d}_t|| - ||\mathbf{d}_s - \mathbf{d}_{ns} || + \varepsilon, 0),
    \end{medsize}
\end{equation}

 \noindent where $||.||$ denotes euclidean distance, and $\varepsilon$ is the margin which is normally set to 1. The original triplet loss implies that the embedding of the manuscript is only guaranteed to be geometrically closed to one target citation. When applying for recommending multiple positive citations, this training objective might be limited.
 
Hence, the three multi-positive triplet objectives are proposed, so that the embedding of the manuscript $\mathbf{d}_s$ should not only be geometrically closed to one target citation $\mathbf{d}_t$, but also be closed to the other positive citations $\{\mathbf{d}_c^n | 1 \leq c \leq n \}$. Meanwhile, the manuscript embedding should be distanced to the $m$ number of negative embeddings $\{ \mathbf{d}_{ns}^m | 1 \leq m \}$.

Three strategies were designed to achieve the objective, which are illustrated in Figure \ref{fig:optim_strategies}:

\begin{itemize}
    \item \textbf{``Target-based'' optimization strategy}: Minimize the distance between $\mathbf{d}_s$ and $\mathbf{d}_t$, and the distances between $\mathbf{d}_t$ and $\{\mathbf{d}_c^n | 1 \leq c \leq n \}$. Hence, when $\mathbf{d}_t$ is found to be similar, embeddings $\{\mathbf{d}_c^n | 1 \leq c \leq n \}$ could also be recommended.
    
    \item \textbf{``Source-based'' optimization strategy}: Minimize the distance between $\mathbf{d}_s$ and $\mathbf{d}_t$, and the distances between $\mathbf{d}_s$ and $\{\mathbf{d}_c^n | 1 \leq c \leq n \}$, so that given $\mathbf{d}_s$, the both of the target and positive citations could be retrieved.
    
    \item \textbf{``Source-target-based'' optimization strategy}: Combining \textbf{``target-based''} and \textbf{``source-based} strategies, it is proposed to minimize the distances between $\mathbf{d}_s$, and both of the target and positive embeddings, and the distances between the target and positive embeddings.

\end{itemize}

\noindent Based on the N-tuplet loss function \cite{DBLP:conf/nips/Sohn16}, we propose three designs of multi-positive triplet objectives following the aforementioned strategies:

\begin{itemize}

    \item Multi-positive target-based triplet (\textbf{\textit{mpt-tgt}}):
    {\scriptsize
    \begin{equation}
            \mathcal{L}_{mpt-tgt} = \log (1 + \sum^{n}_{i=1} \sum^{m}_{j=1}  \exp(||\mathbf{d}_s - \mathbf{d}_t || - ||\mathbf{d}_s - \mathbf{d}_{ns}^j||) \\
            + \exp ( || \mathbf{d}_c^i - \mathbf{d}_t || - || \mathbf{d}_c^i - \mathbf{d}_{ns}^j||))
    \end{equation}
    }

    \item Multi-positive manuscript-based triplet (\textbf{\textit{mpt-src}}):
    {\scriptsize
    \begin{equation}
        % \begin{medsize}
            \mathcal{L}_{mpt-ms} = \log (1 + \sum^{m}_{j=1}  \exp(||\mathbf{d}_s - \mathbf{d}_t || - ||\mathbf{d}_s - \mathbf{d}_{ns}^j||) +  \sum^{m}_{j=1}  \sum^{n}_{i=1} \exp ( || \mathbf{d}_s - \mathbf{d}_c^i || - ||\mathbf{d}_s^i - \mathbf{d}_{ns}^j ||))
        % \end{medsize}
    \end{equation}
    }
    
    \item Multi-positive source-target-based triplet (\textbf{\textit{mpt-src-tgt}}):
    {\scriptsize
        \begin{align}
        % \begin{medsize}
            \nonumber \mathcal{L}_{mpt-src-tgt} = \log (1 + \sum^{m}_{j=1}  \exp(||\mathbf{d}_s - \mathbf{d}_t || - ||\mathbf{d}_s - \mathbf{d}_{ns}^j||) +  \sum^{m}_{j=1} \sum^{n}_{i=1}  \exp ( || \mathbf{d}_s - \mathbf{d}_c^i || \\ - ||\mathbf{d}_s - \mathbf{d}_{ns}^j ||) + \sum^{m}_{j=1}  \sum^{n}_{i=1} \exp(||\mathbf{d}_t - \mathbf{d}_c^i|| - ||\mathbf{d}_t - \mathbf{d}_{ns}^j||))
        % \end{medsize}
        \end{align}
    }
    
\end{itemize}

\vspace{-1em}

\noindent We set the maximum number of positive samples to be 3, since the number of co-citations with greater than 3 pairs is neglectable small from our datasets. \textbf{Mpt-src-tgt} is set to be the default objective for experiments in Section \ref{sec:convention_test}, since it is testified to be the most effective objective according to Section \ref{sec:loss_test}.

\vspace{-1em}

\section{Experiments} \label{sec:experiments}
% This sections illustrates the datasets, implementation details of MP-BERT4CR, and the performance of tests on single and multiple citation recommendations comparing to the baseline models.

\subsection{Dataset}

Four datasets including ACL Anthology (2013 release) and three datasets generated from the DBLP corpus were adapted for experiments. The ACL Anthology corpus includes 20,405 papers with 108,729 citations, whereas the DBLP corpus contains 649,114 papers, with 2,874,303 citations. Three datasets were produced from the DBLP corpus, i.e., DBLP-1, DBLP-2, and DBLP-3, each of which includes 50,000 papers. A ``biased-individuality'' dataset generating strategy was adapted to produce the three DBLP datasets, by which DBLP-2 and DBLP-3 shares 20\% papers (10,000) in-common to evaluate the stability on the performance of the model; whereas the DBLP-1 dataset contains completely different papers to DBLP-1 and DBLP-2. The complete ACL corpus was adapted for the fourth dataset. We split the datasets into train and test sets as listed in Table \ref{tab:dataset}.

% \footnote{Datasets are available at: https://console.cloud.google.com/storage/browser/mp-bert4cr}

% The datasets were divided into a train set for pre-training and fine-tuning the model, and a test set with 20\% of the total amount randomly selected conducting the recommendation tests. The statistics of the datasets are presented in Table \ref{tab:dataset}. ParsCit \cite{DBLP:conf/lrec/CouncillGK08} is applied to parse the in-text citations, so that the algorithms can recognize. ParseLabel \cite{DBLP:journals/ijdls/LuongNK10} was applied to recognize the abstracts, as well as other section headers. The rare words from the vocabulary were compressed to reduce memory consumption by applying byte pair encoding \cite{DBLP:conf/acl/SennrichHB16a} with the adaption of the learned vocabulary table from \citet{DBLP:conf/acl/ZhangWZ19}.

\begin{table}[tb]
\centering
\caption{Statistics of Datasets}
\vspace{-1em}
\label{tab:dataset}
\begin{minipage}{0.9\columnwidth}
\centering
\resizebox{\textwidth}{!}{%
\begin{tabular}{@{}l|l|l|l|l|l@{}}
\toprule
 & \textbf{Paper No. (Total / Train / Test)} & \textbf{Train Cit No.} & \textbf{Test Cit for P$\geq$1} & \textbf{Test Cit for P=1} & \textbf{Test Cit for P$\geq$2} \\ \midrule
\textbf{DBLP-1} & 50,000 / 40,000 / 10,000 & 74,153  & 21,688 & 20,537 & 1,151 \\
\textbf{DBLP-2} & 50,000 / 40,000 / 10,000 & 128,380 & 26,300 & 24,643 & 1,657 \\
\textbf{DBLP-3} & 50,000 / 40,000 / 10,000 & 103,467 & 28,382 & 27,066 & 1,316 \\
\textbf{ACL}    & 20,406 / 16,325 / 4,081  & 31,017  & 21,420 & 15,783 & 188   \\ \bottomrule
\end{tabular}
}
\end{minipage}
\vspace{-2em}
\end{table}
\raggedbottom

%\footnote{{\tiny DBLP-1 and DBLP-2 shares 20\% papers in common to test the consistency of the performances.}}

\subsection{Implementation Details}

MP-BERT4CR was developed based on Fairseq 0.4.0 \cite{ott2019fairseq}, Gensim 2.3.0 \cite{rehurek_lrec}, and Pytorch 1.6.0 \cite{NEURIPS2019_9015}. For the baseline models, Word2Vec and Doc2Vec were implemented using Gensim 2.3.0; HyperDoc2Vec was developed
based on Gensim 2.3.0; DACR was developed based on Pytorch 1.6.0 and Gensim 2.3.0; SciBERT and Specter were implemented using Hugginface 4.2.0 \cite{wolf-etal-2020-transformers}.
%  \footnote{the source code and datasets are available at: https://console.cloud.google.com/storage/browser/mp-bert4cr} 

Adam optimizer \cite{DBLP:journals/corr/KingmaB14} was adapted to optimize MP-BERT4CR with parameters illustrated in Table \ref{tab:params}. The batch sizes are set to be 7 for pre-training, and 1 for fine-tuning. We run 10 iterations of pre-training and 2 iterations of fine-tuning on DBLP-1, DBLP-2, and DBLP-3 datasets; or 50 iterations of pre-training and 5 iterations of fine-tuning on the ACL dataset. The the number of negative samples is set to 4, and maximum number of positive samples is set to 3.

%  Texts with length longer than 30 sentence are sliced into multiple texts for pre-training. 
% introduce baselines
We adapted six baseline modesl: Word2Vec (W2V) \cite{DBLP:journals/corr/abs-1301-3781}, Doc2Vec (D2V) \cite{DBLP:conf/icml/LeM14}, HyperDoc2Vec (HD2V) \cite{DBLP:conf/acl/ShiSZZH18}, DACR \cite{DBLP:conf/coling/ZhangM20}, SciBERT \cite{beltagy-etal-2019-scibert} and Specter \cite{DBLP:conf/acl/CohanFBDW20}. We train the models with default parameters. SciBRET and Specter were fine-tuned with the abstract combined with fundamental contexts (i.e. augmented abstracts), so that the learnt vectors carrying the semantics of the abstract and the citing intents.

\vspace{-1em}

\begin{table}[t]
\caption{Parameters of MP-BERT4CR}
\vspace{-1em}
\label{tab:params}
\centering
\resizebox{0.9\columnwidth}{!}{%
\begin{tabular}{@{}c|c|c|c@{}}
\toprule
\multirow{2}{*}{\textbf{\begin{tabular}[c]{@{}c@{}}Transformer\\ Params (Sent, Doc Encoder, and Decoder)\end{tabular}}} & \textbf{Block No. (L)} & \textbf{Hidden Size (H)} & \textbf{Attention No. (A)} \\ \cmidrule(l){2-4} 
 & 6 & 768 & 12 \\ \midrule
\multirow{6}{*}{\textbf{\begin{tabular}[c]{@{}c@{}}Optimizer\\ Params\end{tabular}}} & \textbf{Update Schedule} & \textbf{Warmup Updates} & \textbf{Update Frequency} \\ \cmidrule(l){2-4} 
 & Inverse Square Root & 1,000 (ACL) / 2,000 (DBLP) & 4 (pretrain) / 8 (finetune) \\ \cmidrule(l){2-4} 
 & \textbf{Warmup Learning Rate} & \textbf{Learning Rate} & \textbf{Weight Decay} \\ \cmidrule(l){2-4} 
 & 1e-7 (pretrain) / 1e-9 (finetune) & 1e-4 (pretrain) / 2e-5 (finetune) & 0.01 \\ \cmidrule(l){2-4} 
 & \textbf{$\beta_1$} & \textbf{$\beta_2$} & \textbf{Dropout} \\ \cmidrule(l){2-4} 
 & 0.9 & 0.999 (pretrain) / 0.98 (finetune) & 0.1 \\ \bottomrule
\end{tabular}
}
\vspace{-1.5em}
\end{table}
\raggedbottom

\begin{table*}[tb]
\caption{Recommendation Scores for Single and Multiple Positive Citations (* $p<0.05$ for paired t test against best baselines)}
\vspace{-1em}
\label{tab:results}
\begin{minipage}{\textwidth}
\centering
\resizebox{0.8\textwidth}{!}{%
\begin{tabular}{@{}llllllllllllllllllllllllll@{}}
\toprule

\multicolumn{1}{l|}{\textbf{Datasets}} & \multicolumn{6}{c|}{\textbf{DBLP-1}} & \multicolumn{6}{c|}{\textbf{DBLP-2}} & \multicolumn{6}{c|}{\textbf{DBLP-3}} & \multicolumn{6}{c}{\textbf{ACL}} \\ \midrule
\multicolumn{1}{l|}{\textbf{No. of Pst. Cits}} & \multicolumn{2}{c}{positive $=$ 1} & \multicolumn{2}{c}{positive $\geq$ 1} & \multicolumn{2}{c|}{positive $\geq$ 2} & \multicolumn{2}{c}{positive $=$ 1} & \multicolumn{2}{c}{positive $\geq$ 1} & \multicolumn{2}{c|}{positive $\geq$ 2} & \multicolumn{2}{c}{positive $=$ 1} & \multicolumn{2}{c}{positive $\geq$ 1} & \multicolumn{2}{c|}{positive $\geq$ 2} & \multicolumn{2}{c}{positive $=$ 1} & \multicolumn{2}{c}{positive $\geq$ 1} & \multicolumn{2}{c}{positive $\geq$ 2}  \\
\multicolumn{1}{l|}{\textbf{Metrics}} & Recall & MAP & Recall & MAP & Recall & \multicolumn{1}{l|}{MAP} & Recall & MAP & Recall & MAP & Recall & \multicolumn{1}{l|}{MAP} & Recall & MAP & Recall & MAP & Recall & \multicolumn{1}{l|}{MAP} & Recall & MAP & Recall & MAP & Recall & MAP \\ \midrule
\multicolumn{1}{l|}{\textbf{W2V}} & 13.26 & 6.29 & 13.26 & 6.29 & 13.48 & \multicolumn{1}{l|}{7.10} & 19.25 & 10.25 & 19.35 & 10.38 & 20.81 & \multicolumn{1}{l|}{12.41} & 17.76 & 9.09 & 17.76 & 9.09 & 17.77 & \multicolumn{1}{l|}{9.15} & 17.16 & 8.2 & 17.13 & 8.25 & 14.37 & 10.37  \\
\multicolumn{1}{l|}{\textbf{D2V}} & 2.07 & 0.48 & 2.07 & 0.48 & 2.17 & \multicolumn{1}{l|}{0.68} & 1.27 & 0.29 & 1.27 & 0.30 & 1.26 & \multicolumn{1}{l|}{0.38} & 2.44 & 0.78 & 2.45 & 0.79 & 2.48 & \multicolumn{1}{l|}{0.92} & 2.22 & 5.73 & 2.23 & 5.75 & 2.69 & 0.74 \\
\multicolumn{1}{l|}{\textbf{HD2V}} & 34.15 & 17.50 & 34.15 & 17.50 & 41.01 & \multicolumn{1}{l|}{22.49} & 40.12 & 21.56 & 40.53 & 21.86 & 46.59 & \multicolumn{1}{l|}{26.22} & 40.64 & 21.64 & 40.67 & 21.64 & 41.35 & \multicolumn{1}{l|}{21.55} & 30.91 & 16.40 & 30.83 & 16.34 & 23.63 & 11.97 \\
% \multicolumn{1}{l|}{\textbf{DC2V}} & 35.36 & 17.58 & 35.36 & 17.58 & 38.86 & \multicolumn{1}{l|}{21.06} & 39.99 & 20.18 & 40.20 & 20.38 & 43.33 & 23.28 \\
\multicolumn{1}{l|}{\textbf{DACR}} & 22.42 & 10.34 & 22.42 & 10.34 & 25.06 & \multicolumn{1}{l|}{11.94} & 27.42 & 14.07 & 27.79 & 14.36 & 33.21 & \multicolumn{1}{l|}{18.51} & 27.15 & 13.48 & 27.14 & 13.45 & 26.89 & \multicolumn{1}{l|}{12.92} & 24.31 & 12.21 & 24.25 & 12.19 & 19.49 & 10.93 \\
\multicolumn{1}{l|}{\textbf{SciBERT}} & 8.95 & 4.03 & 8.95 & 4.03 & 10.23 & \multicolumn{1}{l|}{5.23} & 7.79 & 3.40 & 7.87 & 3.46 & 9.06 & \multicolumn{1}{l|}{4.39} & 10.42 & 4.64 & 10.51 & 4.70 & 12.45 & \multicolumn{1}{l|}{6.01} & 0.16 & 0.03 & 0.16 & 0.03 & 0.21 & 0.06 \\
\multicolumn{1}{l|}{\textbf{Specter}} & 4.51 & 2.26 & 4.51 & 2.26 & 5.42 & \multicolumn{1}{l|}{3.72} & 3.42 & 1.67 & 3.45 & 1.74 & 3.89 & \multicolumn{1}{l|}{2.71} & 4.51 & 2.26 & 4.51 & 2.26 & 5.42 & \multicolumn{1}{l|}{3.72} & 3.42 & 1.67 & 3.45 & 1.74 & 3.89 & 2.71 \\ \midrule
\multicolumn{1}{l|}{\textbf{MB4R$_{no\_dynamic}$}} & 38.97 & 18.98 & 39.42 & 19.33 & 47.29 & \multicolumn{1}{l|}{25.48} & 39.51 & 18.71 & 40.00 & 19.03 & 47.21 & \multicolumn{1}{l|}{23.67} & 43.35 & 21.69 & 43.77 & 21.96 & 52.30 & \multicolumn{1}{l|}{27.54} & 30.47 & 14.31 & 30.50 & 14.34 & 32.78 & 17.09 \\
\multicolumn{1}{l|}{\textbf{MB4R$_{triplet}$}} & 40.07 & 18.08 & 40.34 & 18.33 & 45.04 & \multicolumn{1}{l|}{22.79} & 40.44 & 19.51 & 40.79 & 20.01 & 48.24 & \multicolumn{1}{l|}{27.73} & 47.52 & 22.61 & 48.05 & 22.98 & 59.02 & \multicolumn{1}{l|}{30.24} & 30.82 & 13.23 & 30.92 & 13.29 & 39.55 & 18.30 \\

\multicolumn{1}{l|}{\textbf{MB4R$_{mpt-src-tgt}$}} & \textbf{44.81*} & \textbf{21.45*} & \textbf{45.21*} & \textbf{21.77*} & \textbf{52.49*} & \multicolumn{1}{l|}{\textbf{27.50*}} & \textbf{47.25*} & \textbf{23.41*} & \textbf{47.66*} & \textbf{23.76*} & \textbf{53.69*} & \multicolumn{1}{l|}{\textbf{29.10*}} & \textbf{48.86*} & \textbf{23.42*} & \textbf{49.38*} & \textbf{23.73*} & \textbf{60.05*} & \multicolumn{1}{l|}{\textbf{30.60*}} & \textbf{36.13*} & \textbf{16.90*} & \textbf{36.18*} & \textbf{16.93*} & \textbf{40.61*} & \textbf{19.89*} \\ \midrule

\end{tabular}
}
\end{minipage}
\vspace{-1.5em}
\end{table*}
\raggedbottom

\subsection{Recommendation Methodology}

\label{sec:convention_test}
 We present the methods for generating recommendations for MP-BERT4CR (MB4R thereafter) and the baselines. Recall and MAP at top 10 recommendations were reported for analyses.

For MB4R, we firstly conduct dynamic sampling for the papers in the test and train set. The sampled manuscript and citation contexts are then encoded via the fine-tuned manuscript encoder or citation encoder to get \textbf{query vectors} and \textbf{citation vectors}. Recommendations were selected as the top 10 citations by cosine similarities.

For W2V, D2V, HD2V, and DACR, we use the same method from the original studies \cite{DBLP:conf/acl/ShiSZZH18,DBLP:conf/coling/ZhangM20}. The \textbf{query vectors} are computed by averaging the word vectors of the local context. The \textbf{citation vectors} are computed via different methods. For \textbf{W2V}: we use the input embedding of the citation ids as \textbf{citation vectors}; for \textbf{D2V}: we use the trained model to infer \textbf{citation vectors} from the content words; and for\textbf{HD2V,DACR}: use the output embedding of the citation ids as \textbf{citation vectors}. For \textbf{SciBERT} and \textbf{Specter}, we generate \textbf{query vectors} by using the fundamental context sampled. The \textbf{citation vectors} are produced by encoding and sum-pooling the augmented abstracts. Recommendations  are made by cosine similarities.

\begin{table}[t]
\centering
\caption{Comparison on Multi-Positive Triplet Objectives with Conventional Triplet on DBLP-1}
\vspace{-1em}
\label{tab:losses}
\resizebox{0.8\columnwidth}{!}{%

\begin{tabular}{@{}l|llllll@{}}
\toprule
\textbf{No. of Positive Cits} & \multicolumn{2}{c}{\textbf{positive = 1}} & \multicolumn{2}{c}{\textbf{positive $\mathbf{\geq}$ 1}} & \multicolumn{2}{c}{\textbf{positive $\geq$ 2}} \\ \midrule
\textbf{Metrics} & \textbf{Recall} & \textbf{MAP} & \textbf{Recall} & \textbf{MAP} & \textbf{Recall} & \textbf{MAP} \\ \midrule
\textbf{Best Baseline} & 34.15 & 17.50 & 34.15 & 17.50 & 41.01 & 22.49 \\
\textbf{MB4R$_{triplet}$} & 40.07 & 18.08 & 40.34 & 18.33 & 45.04 & 22.79 \\
\textbf{MB4R$_{Mpt-src}$} & 41.86 & 19.99 & 42.23 & 20.22 & 48.82 & 24.26 \\
\textbf{MB4R$_{Mpt-tgt}$} & \textbf{44.85} & \textbf{21.69} & 45.10 & \textbf{21.94} & 49.68 & 26.53 \\
\textbf{MB4R$_{Mpt-src-tgt}$(default)} & 44.81 & 21.45 & \textbf{45.21} & \textbf21.77 & \textbf{52.49} & \textbf{27.50} \\ \bottomrule
\end{tabular}
}
\vspace{-1.5em}
\end{table}

\raggedbottom

\subsection{Analysis on Citation Recommendations}

\subsubsection{Single Positive}
Two points could be drawn from Table \ref{tab:results} ($positive=1$ cases). First, MB4R outperformed all the baseline models across all the datasets and metrics by significant margins, from which MB4R$_{triplet}$'s superiority compared with baselines testified the effectiveness of the proposed neural architecture, and MB4R$_{mpt-src-tgt}$ further testified the effectiveness of the proposed multi-positive objectives. Second, the multi-positive objective function is not only helpful to identify multiple ground-truth citations, the single positive performances are also improved when comparing MB4R$_{mpt-src-tgt}$ to MB4R$_{triplet}$. In addition, owing to the more hierarchical transformer and dynamic sampling strategy, MB4R outperformed Sci-BERT and Specter.

\subsubsection{Multiple Positive }
% \subsection{Analysis on Multiple-Positive Citation Recommendations}

According to the scores with multiple ground-truth citations (cases of $positive \geq 1$ and $positive \geq 2$) in Table \ref{tab:results}, the scores for multiple positive recommendations from MB4R$_{mpt-src-tgt}$ remained effective comparing with the baselines and MB4R$_{triplet}$. However, as the $positive \geq 2$ test samples are much less than the test samples for $positive \geq 1$, so the ability of \textit{mpt} might not be fully demonstrated, especially on DBLP-3 and ACL datasets. We will produce datasets with a higher number of positive samples in the later stage for further tests.

\vspace{-1em}

\subsection{Comparisons on Different Designs of Multi-Positive Objectives}
\label{sec:loss_test}

We compare the three designs of multi-positive objectives in Table \ref{tab:losses}. First, all the three proposed multi-positive objectives (MB4R$_{mpt-src}$, MB4R$_{mpt-tgt}$, and MB4R$_{mpt_src-tgt}$) outperformed the best baseline (HD2V), and the original triplet objective (MB4R$_{triplet}$). Second, the two target-based strategies MB4R$_{mpt-tgt}$ and MB4R$_{mpt_src-tgt}$ are superior than the source-based strategy (MB4R$_{mpt-src}$), which means that the distances between the ground-truth candidate and other positive candidates play the central role for multiple retrieving. Third, MB4R$_{mpt_src-tgt}$ produced superior performances comparing with MB4R$_{mpt-tgt}$ for case \textit{positive $\geq 2$}, but very close performances for cases \textit{positive $= 1$} and \textit{positive $\geq 1$}, and hence it was set as the default. We also tested MB4R with or without dynamic sampling for an ablation analysis, it is found that the dynamic sampling mechanism can significantly improve the performances, by comparing the resutls from $MB4R_{mpt-src-tgt}$ and $MB4R_{no_dynamic}$.

\vspace{-1em}

\section{Conclusion}
\vspace{-0.5em}
In this study, we proposed MP-BERT4CR, a content modeling for multi-positive citation recommendations. It has the following advantages: first, it comes with a series of multiple positive objectives to optimize the model for multi-positive recommendation; second, it can effectively recommend co-citations by leveraging the historical patterns; Third, the proposed dynamic context sampling strategy empowers the citation embeddings to carry content semantics, to further improve the performances.

%  Experimental results verified the effectiveness on single and multiple positive recommendations, retrieving high and low frequent co-citation pairs, and full list of co-citation items. 

\vspace{-1em}
\begin{acks}
\vspace{-0.5em}
This research has been supported in part by JSPS KAKENHI under Grant Number 19H04116.
\end{acks}
\vspace{-1em}


\bibliographystyle{ACM-Reference-Format}
\bibliography{references}


\end{document}